\documentclass[aps,prl,preprint,groupedaddress,graphics]{revtex4}

\usepackage{graphicx}
\usepackage{dcolumn}
\usepackage{amsmath}
\usepackage{bm}

\begin{document}

\title {On the conflict between local realism and classical physics}
\author{S. Savasta, O. Di Stefano,  R. Girlanda}
\affiliation{Dipartimento di Fisica della Materia e Tecnologie
Fisiche Avanzate, Universit\`{a} di Messina Salita Sperone 31,
I-98166 Messina, Italy}

\begin{abstract}

In contrast to the intuitively plausible assumption of local
realism, entangled particles, even when isolated, are not allowed
to possess definite properties in their own right , as
quantitatively expressed by  violations of Bell's inequalities \cite{Rep}.
Even as entanglement is now a key feature of quantum information
and communication technology \cite{bendiv,Nielsen}, it remains the  most puzzling
feature of quantum mechanics \cite{Bellbook} and its conceptual foundation is
still widely debated. Here we demonstrate that  physical systems
providing dicotomic outcomes are not able to guarantee both the rotation
properties of physical quantities and local realism. This result
opens the way to a new formulation of quantum mechanics based on
only two elementary physical principles replacing the 
abstract mathematical axiomes of the present theory.
According to this formulation, the coexistence of discrete
outcomes with the classical continuous transformation properties
of physical quantities under coordinate transformation inevitably
implies quantum probabilities. These results, provide a simple physical
explanation to the most debated quantum features and put into
question the existence of physical quantities displaying
continuous outcomes in agreement with approaches that attempt to integrate quantum theory with general relativity \cite{Camelia,Dreyer,Baez,Rovellibook}.

\end{abstract}

\maketitle

\newpage

Einstein  reduced the abstract mathematical structure of the
Lorentz transformations to two simple physical principles expressible in
common language.
Although the theory of
special relativity leads to surprising and in part even
counterintuitive consequences, thanks to the existence of these
physical principles we do not have a significant debate on the
interpretation of the theory of special relativity. The
formulation of quantum mechanics, to the contrary, is based on a
number of rather abstract, axioms. The motivations for the
postulates is not always clear, and they appeared surprising even
to the founding fathers of quantum theory. This absence of
elementary physical principles together with the counter-intuitive
consequences of the postulates  determined a relentless broad
discussion about the interpretation of the theory \cite{Schlosshauer}, despite its
success.
Albert Einstein disliked the loss of determinism in measurement.
He held that quantum mechanics must be incomplete, and produced a
series of objections to the theory. The most famous of these was
the Einstein Podolsky Rosen (EPR) paradox \cite{EPR}. The EPR paradox was
advanced as an argument that quantum mechanics could not be a
complete theory but should be supplemented by additional (hidden)
variables to restore causality and locality. Bell's theorem put
forward the conflict between  quantum mechanics  and hidden
variables predictions \cite{Bell}. Later experimental verification of
violations of Bell inequalities \cite{Aspect,Rowe} disproved local realism and hidden
variable theories, showing that nature is intrinsically not
deterministic as predicted by quantum theory. Nevertheless John
Bell himself was profoundly unsatisfied with  the axiomatic
formulation of quantum theory \cite{Bellbook}. According to him ordinary quantum
mechanics  is  just fine for all practical
purposes (abbreviated, rather disparagingly, as
``FAPP''). Recently the field of quantum information theory
opened up and expanded rapidly. Quantum entanglement began to be
seen not only as a puzzle, but also as a resource which can yield
new physical effects and techniques \cite{Nielsen}.
In turn  ideas from these fields are beginning to yield new
insight into the foundations of quantum physics, suggesting that
information should play an essential role in the foundations of
any scientific description of Nature \cite{Found,SGBattista,BruZei}. 
Besides, classical physics  appears to play a not dismissable role.
Despite quantum theory provides probably the most striking departure from
classical ideas of reality, it requires for its formulation classical
concepts \cite{Landaubook}.
Here we explore this deep connection by adopting a
conservative approach, trying to introduce into classical physics
the smallest modifcations able to describe experimental evidences
of light quanta.

We  start our analysis pointing out a not new and
rather obvious contact point between classical and quantum
physics. The polarization status of a monochromatic light beam
can be represented by a  unit vector ${\bf J}$ on the Poincar\'e
sphere. The outcome of polarization measurements along a direction
${\bf n}$ on the Poincar\'e sphere can be described by ${\bf J}
\cdot {\bf n}$. Let us consider an  horizontally-polarized light
beam. Of course further measurements of polarization in the H/V
(horizontal/vertical) basis  will say that all the light will
follow the H direction (${\bf J} \equiv (0,0,1)$). We consider now
measurements of linear polarization along directions $H'/V'$
(along the ${\bf x}$ axis of the Poincar\'e sphere) rotated by
$45^\circ$ with respect to the horizontal axis . According to
classical physics, the light beam will be splitted into two  beams
of equal intensities $I_{H'}= I_{V'}$, giving rise to  ${\bf J}
\cdot {\bf x}= (I_{H'}-I_{V'})/(I_{H'}+I_{V'})=0$. If the beam is
progressively attenuated before passing the polarizing beam
splitters, nature tells us that the splitting into two beams
cannot be continued indefinitely, we reach a situation where a
single light quantum will follow one direction only.
If the photon cannot be divided which are the new rules? Which
direction will it follow? Actually we will see photons  choosing
either path $H'$ or $V'$  randomly. If we use the experimental
outcomes (e.g. the number of photons $n_{H'}$ and $n_{V'}$
detected by the two photon-counters after repeated
$N=n_{H'}+n_{V'}$ events), the mean value of the Stokes parameter
$(n_{H'}-n_{V'})/N$  will approach zero when increasing $N$:
$\left< {\bf J} \cdot {\bf x} \right> = \lim_{N \to \infty}
(n_{H'}-n_{V'})/N= 0$, i.e. it will be zero in the mean. Thus
$\left< {\bf J} \cdot {\bf x} \right>$ follows classical physics.
So in a situation where physicists were forced to abandon the ship
of classical physics, this contact point can be viewed as  a  life
raft. {\em Once we accept the presence of discrete outcomes, the
emerging of quantum probabilities can be viewed as the only means
that nature has to follow the transformation rules of classical
physics}.

Let us now consider a system of two particles with zero total
angular momentum. According to classical physics this imply that
if we measure projections of total angular momentum along an
arbitrary axis ${\bf \hat n}$, we will find zero, so ${\bf J}
\cdot {\bf \hat n}$ and $|{\bf J}|$ will be zero. We may assume
that total angular momentum is conserved even if the two particles
are well separated and have ceased to interact. If we also assume
that measurements of angular momentum along a given axis take
values only in $\left\{ \pm 1 \right\}$, the only possibility for
the existence of two particles with zero total angular momentum is
that they display perfect anticorrelation for measurements along
arbitrary axes, i.e. if  particle 1 provides the outcome
$v_{1}({\bf n})= \pm1$, the other will provide opposite values so
that $v_1({\bf n})v_2({\bf n})= -1$.
Just for continuity with the previous example, let us consider
light quanta. Each observer has his own detection apparatus PBS
${\bf n}$ able to perform  polarization measurements along  a
direction {\bf n} on the Poincar\'e sphere.
The CHSH inequality \cite{CHSH},  on which many experimental tests of Einstein {\em
local realism} have been performed,  is part of the large set of
inequalities known generally as Bell inequalities. It applies to a
situation in which observer 1 which receives particle 1 can choose
to use either PBS ${\bf n}$ or PBS ${\bf n}'$. Analogously
observer 2 can use either PBS ${\bf m}$ or PBS ${\bf m}'$.
The observables $v_1$ and $v_2$ take values only in $\left\{ \pm 1
\right\}$ and may be functions of hidden random variables. The
inequality can be expressed as $\left| \left< B \right> \right|
\leq 2$, where
$
    B \equiv \left[ v_1({\bf n})+ v_1({\bf n}')\right] v_2({\bf m}) + \left[ v_1({\bf n})- v_1({\bf n}')\right] v_2({\bf m}')= \pm 2
$.
It  poses a limit to the degree of correlation
permitted by a theory assuming dicotomic outcomes and local
realism.
We now ask if  the mean values of these variables satisfy (at
least statistically) the laws of classical physics as the Stokes
parameter examined before. To this aim we consider a system with
zero total angular momentum.
We start addressing the situation where  the two observers perform
measurements along the same axis ${\bf n}$. Each run of the
experiment will give rise to $v_1({\bf n})v_2({\bf n})= -1$. What
happens if observer 2 uses a differently oriented apparatus PBS
${\bf m}$? According to the  transformation properties of vectors
under rotations, from $v_1({\bf n})v_2({\bf n})= -1$, it results
\cite{nota} $v_1({\bf n})v_2({\bf m})= -{\bf n} \cdot {\bf m}$. Of
course theories with discrete outcomes cannot give this  result
that is not in $\left\{ \pm 1 \right\}$. Driven by the previous
example we may require that this result holds in the mean: $\left<
v_1({\bf n})v_2({\bf m})\right>= -{\bf n} \cdot {\bf m}$. Now we
can check this result against the CHSH inequality. We consider the
case where ${\bf n}'$, ${\bf m}$, ${\bf n}$, ${\bf m}'$ are
coplanar and separated by successive $45^\circ$; we obtain $
\left| \left< B \right> \right| = 2\sqrt{2} $ in clear violation
of the CHSH inequality. Thus the combination of discrete outcomes
and classical physics (in a statistical meaning) gives rise to a
violation of the CHSH inequality. The following theorem holds:
{\em Local realism, the transformation properties of vectors in
classical physics (followed at least in the mean), and discrete
outcomes cannot hold all together.} We also observe that the
obtained result $\left< v_1({\bf n})v_2({\bf m})\right>= -{\bf n}
\cdot {\bf m}$ coincides with  predictions of ordinary quantum
mechanics. Just starting from discrete outcomes and assuming
transformation properties of vectors on average we have obtained
the  violations of the CHSH predicted by quantum theory. It is
worth noting that the present result has been obtained without
using  Hilbert spaces , tensor products,  Hermitian operators,
Pauli matrixes.
According to this analysis violations of Bell inequalities (usually regarded
as the most striking departure of quantum mechanics from classical
physics) is the only possibility for a physical system with zero
total angular momentum and with dicotomic outcomes to follow the
transformation properties of vectors. From this point of view
local realistic theories are more nonclassical than quantum
mechanics as they do not follow even in the mean the
transformation rules of classical physics.

The experimental tests of the CHSH inequalities  confirm that
nature choose to follow the laws of classical physics as far as
allowed (in the mean) by the presence of discrete outcomes, giving
up local realism \cite{Aspect,Rowe}. Here local realism seems to play the role of
absolute time in special relativity. The theorem and this analogy
suggest a  formulation of  quantum theory by means of only two
general principles. A first principle accounts for many
experimental evidences that  microscopic systems if asked provide
discrete outcomes. This  may appear  reasonable. For example, it
seems to us  reasonable that a light beam cannot  be subdivided
indefinitely. Feynman wrote: `` It always bothers me that,
according to the laws as we understand them today, it takes a
computing machine an infinite number of logical operations to
figure out what goes on in no matter how tiny a region of space,
and no matter how tiny a region of time \dots Why should it take an infinite amount of
logic to figure out what one tiny piece of space/time is going to
do?''. The only possibility to describe what happens in a finite
region of space with a finite amount of information is the
assumption of discrete outcomes. Of course the concept of
discretization does not produce automatically quantum physics;
classical information theory, for example, even if based on
discretization is not quantum. The theorem here presented
indicates that  continuous transformation properties of classical
physical quantities under coordinate transformation may be the
additional ingredient giving rise to quantum physics.

In 1999 Anton Zeilinger proposed the following  information-based
foundational principle for quantum mechanics \cite{Found}: ``{\em The most
elementary system carries just one bit of information}''. As
remarked by Zeilinger this principle is basic and elementary
enough that it actually can serve as a foundational principle for
quantum mechanics. Some of the essential features of quantum
mechanics as the irreducible randomness of individual events,
quantum complementary and quantum entanglement, arise in a natural
way from it. Of course discrete outcomes would be a direct
consequence of this principle. We adopt this principle as the
first principle for quantum physics. As remarked above, the only
possibility for a system with discrete outcomes to follow
continuous classical transformations is to adhere to them in the
mean. The second principle has to contain somewhat this adherence.
We propose: {\em Physical systems satisfy the  laws of classical
physics in the mean when the first principle prevents a
deterministic adherence}. This is a simple conservative principle preventing any conflict between the laws of classical physics (including relativity theories).
After accepting the two principles, the theorem becomes  a direct
consequence of them. Hence the two principles imply quantum
entanglement and consequent violations of the CHSH inequality. In
the following we will see as the principles also allow a direct
derivations of other relevant and counterintuitive quantum
results.
We start considering what in ordinary quantum theory is the
projection postulate.  If we perform  a first measurement on a
physical system, we obtain one specific discrete outcome (as
prescribed by the first principle). If we repeat again the same
measurement (without affecting the system in the meanwhile between
the two measurements), we will obtain deterministically the same
discrete outcome obtained in the first measurement as happens in
classical physics (according to the second principle). This is
consequence of the fact that in this case the classical result
doesn't conflict with discrete outcomes. As an example we may
refer once again to the polarization of classical light. If we
start with e.g. a $V$-polarized beam and send it to one or more
PBS in the H/V basis, the polarization state will not be modified
and the beam will always follow the V path.

We now begin from elementary  systems and we face with the
problem of describing classical properties, as the transformation
of vectors under rotations, with only two possible outcomes $\pm
a$ (1 bit).
Specifically we assume that we obtained the outcome $+a$. Thus the
system is oriented along the ${\bf \hat n}$ axis \cite{nota}  with
positive direction: ${\bf J}= {\bf \hat n}$. If we perform a
polarization measurement along the ${\bf \hat n}$ axis,  according
to classical physics we obtain ${\bf J} \cdot {\bf \hat n}=a$.
This result is in $\{ \pm a\}$  hence, according to the second
principle, it is also the quantum result.
What happens if we choose  a different detection axis ${\bf \hat
m}$?
According to classical physics we obtain \cite{nota} ${\bf J}
\cdot {\bf \hat m}=a{\bf \hat n} \cdot {\bf \hat m} = a\cos
\theta$, being $\theta$ the angle determined by the two axis. This
result is not in $\{ \pm a \}$  hence, according to the second
principle, the quantum result follows the classical one only in
the mean: $\left< {\bf J} \cdot {\bf \hat m} \right>_{{\bf \hat
n+}} =a \cos \theta$ and the expectation value can be expressed in
terms of probabilities $P({{\bf \hat n}\, \pm},{{\bf \hat m}\,
\pm})$ as $\left< {\bf J} \cdot {\bf \hat m} \right>_{{\bf \hat
n+}}= a\left< {\bf \hat n} \cdot {\bf \hat m} \right> \equiv
aP({\bf \hat n}+:{\bf \hat m}+) - aP({\bf \hat n}+ : {\bf \hat
m}-)$. As a consequence we have the following equation
\begin{equation}
 a P({\bf \hat n}+:{\bf \hat m}+) - aP({\bf \hat n}+:{\bf \hat m}-) = a\cos \theta\, .
\end{equation}
By using that probabilities must sum to unity: $P({\bf \hat n} \pm
:{\bf \hat m}+) + P({\bf \hat n}\pm :{\bf \hat m}-) = 1$,
 we can readily derive quantum  probabilities
\begin{eqnarray}
P({\bf \hat n}+:{\bf \hat m}+) = \cos^2 (\theta/2)\nonumber\\
P({\bf \hat n}+:{\bf \hat m}-) = \sin^2 (\theta/2)\, .
\end{eqnarray}
Analogously, starting from the outcome $-a$, we would obtain
$P({\bf \hat n}- : {\bf \hat m}\pm) =P({\bf \hat n}+ : {\bf \hat
m}\mp)$.
This result shows how the probabilistic nature of quantum theory
descends directly from the two principles. In particular we
derived the correct quantum probabilities prescribed by quantum
mechanics starting from only two principles. We notice that, as
expected, probabilities depend only on the relative angle between
the two involved directions, hence exchanging the two directions
does not modify probabilities. Also exchanging the initial and
final outcomes does not modify probabilities: $P({\bf \hat
n}+:{\bf \hat m}-)= P({\bf \hat n}-:{\bf \hat m}+)$. We interpret
this finding as a sort of probabilistic reversibility, that is
what survive of classical reversibility after the effects of the
two principles. Once we obtain probabilities for a measurement of
{\bf J} along a given axis we are able to obtain information about
its variance by deriving
\begin{equation}
\left< ({\bf J} \cdot {\bf \hat m})^2 \right>_{{\bf \hat n}\beta}
= \sum_{\alpha=\pm 1} (\alpha a)^2 P({\bf \hat n}\beta:{\bf \hat
m}\alpha)\,.
\end{equation}
From probability conservation the above expression is a constant
equal to $a^2$. As a consequence the expectation value of the
square modulus of {\bf J} results to be a scalar according to the
second principle and in agreement with quantum mechanics: $\left<
{\bf J}^2 \right>=3a^2$. The obtained results coincide with
corresponding results obtained by using 2D Hilbert spaces and the
theory of angular momentun of ordinary quantum theory. So as far
as we calculate transition probabilities for vectors with two
outcomes, it can be useful to use 2D Hilbert spaces just for all
practical purposes. For example  transition probabilities
can be obtained in the usual mathematically elegant way as
$P({\bf \hat m} j : {\bf \hat n} k) = |\left< {\bf \hat n} k |
{\bf \hat m} j \right>|^2$, being $\left|{\bf \hat m} j \right>$ a normalized vector in a 2D Hilbert space.

A direct consequence of the first principle is that  elementary
composite systems may be tought as constituted by two elementary
systems and, hence, carry two bits.
We consider  measurements along a
given ${\bf \hat n}$ axis on such systems. According to the first principle there
are three different possibilities: $\pm 2a$ when both the values
of the two elementary systems are equal and $0$ when they assume
opposite values. We start considering the outcome $+2a$
corresponding to the couple of states
$\{{\bf \hat n}+ \}_1|\{ {\bf \hat n}+\}_2$. What happens if we choose  a
different detection axis ${\bf \hat m}$? We derive the
probabilities $P({\bf \hat n}2 : {\bf \hat m}k)$ (with $k=\pm2,0$)
of obtaining $\pm2a,0$ when performing measurements along ${\bf
\hat m}$. The probability of obtaining $\pm 2a$ is given by the
probability that both elementary systems give $\pm a$: $P({\bf \hat
n}2 : {\bf \hat m}\pm 2)=P_1({\bf \hat n}+ : {\bf \hat m}\pm)P_2({\bf
\hat n}+ : {\bf \hat m}\pm) = (1\pm\cos \theta)^2/4$. 
The outcome $0$ arises if the first elementary outcome assumes  values
$\pm a$ and the second outcome gives opposite values $\mp a$. Hence,
$P({\bf \hat n}2 : {\bf \hat m}0)=P_1({\bf \hat n}+ : {\bf \hat
m}+)P_2({\bf \hat n}+ : {\bf \hat m}-)+P_1({\bf \hat n}+ : {\bf
\hat m}-)P_2({\bf \hat n}+ : {\bf \hat m}+)$ and we obtain
\begin{equation}
    P({\bf \hat n}2 : {\bf \hat m}0)=\frac{1}{2}(1-\cos^2 \theta).
\end{equation}
This result can also be obtained by exploting that probabilities
must sum to one: $\sum_k P({\bf \hat n}2 : {\bf \hat m} k)= 1$
(with $k = 0, \pm2$).
Analogous results can be derived for $P({\bf \hat n}-2 : {\bf \hat
m}k)$. Oerived probabilities satisfy probabilistic
reversibility: $P({\bf \hat n}-2 : {\bf \hat m}2) = P({\bf \hat
n}2 : {\bf \hat m}-2)$.
The consistency of this approach can be checked by using the
derived probabilities to  inspect the rotation properties of  the
expectation values $ \left< {\bf \hat n} \pm 2 \right| \bm{ J}
\left|{\bf \hat n}\pm 2 \right>$ and $\left<{\bf \hat n} \pm
2|{\bf J}^2 | {\bf \hat n} \pm 2\right>$. They can be readily
calculated by using the obtained probabilities and it turns out
that the first is a vector and the second a scalar (equal to
$8a^2$) consistently with the second principle and with QM.
In order to complete we have to derive $P({\bf \hat n}0: {\bf \hat
m}k)$. The physical state ${\bf \hat n}0$ brings some complication
due to the fact that outcome zero can be obtained with two
distinct possibilities: zero is obtained when the second system
gives outcomes which are opposite to those of the first. In
particular this outcome is produced by the two couples of
elementary systems: $|{n}\pm \rangle_1|{n}\mp\rangle_2$. Hence we
do not know to which of the two couples a specific realization of
the outcome zero corresponds. The most simple idea is to think to
an equal mixture of the two systems, e.g. when once obtains zero,
ones obtained or $|{n}+ \rangle_1|{n}-\rangle_2$ either $|{n}-
\rangle_1|{n}+\rangle_2$ each with a $50\%$ of probability.
Following this procedure we obtain $P({\bf \hat n}0: {\bf \hat
m}2) = \left[P_1({\bf \hat n}+ : {\bf \hat m}+)P_2({\bf \hat n}- :
{\bf \hat m}+)+P_1({\bf \hat n}- : {\bf \hat m}+)P_2({\bf \hat n}+
: {\bf \hat m}+) \right]/2$. Hence $P({\bf \hat n}0: {\bf \hat
m}2) = P({\bf \hat n}2: {\bf \hat m}0)/2$ in contrast to previous
cases. According to this result transitions $\pm 2a \to 0$ would
be more favorite than transitions $0 \to \pm 2a$. This can be
understood as a consequence of the hidden information on which of
the two elementary systems gives $+a$. This derivation attributes
elements of physical reality to the two elementary subsystems and
opens the possibility of some hidden variables associated to the
zero outcome.
By using this proposed probability $P({\bf \hat n}0: {\bf \hat
m}2)= \frac{1}{4}(1-\cos^2 \theta)$, we would obtain $\left<{\bf
\hat n} 0|{\bf J}^2 | {\bf \hat n} 0 \right> = 4a^2 \neq\left<{\bf
\hat n} \pm 2|{\bf J}^2 | {\bf \hat n} \pm 2 \right>$.
This implies that a rotation can produce a change in the value of the scalar $\left< {\bf
J^2} \right>$ in contrast with the second principle. Again
elements of reality, discrete outcomes, and rotation properties of
vectors conflict. The correct transition
probability can be derived regarding the components $\left< J^2_i \right>$ as the diagonal elements of a symmetric rank 2
tensor $M_{ij}$.
We start from the state ${\bf \hat z} 0$. We know that
$M_{33}= 0$. We also know that Tr{\bf M} = $\left< {\bf J^2}
\right> = 8a^2$.  Symmetry implies $M_{11}= M_{22}$ and $M_{ij}=
0$ for $i \neq j$. According to the transformation rules for
tensors, we obtain $M^\prime_{33}=4a^2\sin^2 \theta$ where
$\theta$ is the angle between the ${\bf \hat z}$  and the new
measurement axis. According to our principles this result has to
hold statistically, hence we have $P({\bf \hat n}0: {\bf \hat
m}2)= P({\bf \hat n}2: {\bf \hat m}0) = \frac{1}{2}\sin^2 \theta$.
Thus also in this case reversibility holds. This   result shows
that we cannot attribute elements of physical reality to the two
elementary subsystems. They cannot be interpreted as an equal
mixture of the two systems but as an indistinguishable combination
of them in such a way that the probability $P({\bf \hat n}0: {\bf
\hat m}2)$ results as the sum of two contributions: $P_1({\bf \hat
n}+ : {\bf \hat m}+)P_2({\bf \hat n}- : {\bf \hat m}+)+P_1({\bf
\hat n}- : {\bf \hat m}+)P_2({\bf \hat n}+ : {\bf \hat m}+)$.
Indistinguishability gives rise to an additional  factor two in
the transition probability in analogy with constructive
interference arising from coherent waves. It is remarkable that wave-like effects comes out by
imposing only the rotation properties of vectors in presence of discrete outcomes. Indeed 
obtained results agree with the
theory of angular momentun of ordinary quantum theory.
Also in this case probabilities can be expressed as :
$P({\bf \hat m} j : {\bf \hat n} k) = |\left< {\bf \hat n} k |
{\bf \hat m} j \right>|^2$.
Dealing with the Hilbert space, we notice that rotations mix the three states $| {\bf
\hat n},k \rangle$ spanning a 3D subspace of the 4D space. The
remaining 1D subspace is spanned by the vector $| \psi^- \rangle =
| {\bf \hat m},0 \rangle=\frac{1}{\sqrt 2}(| {\bf \hat m} +, {\bf
\hat m} - \rangle-| {\bf \hat m} -, {\bf \hat m} + \rangle)$ that,
of course, is orthogonal to the other three vectors. This state
corresponds to the outcome zero. It is invariant if expressed in a
different basis (as it is well known), thus produces only zero
outcomes even if we change measurement axis. It gives $\langle
\psi^- | {\bf J} \cdot {\bf \hat m}| \psi^- \rangle=0$ and
$\langle \psi^- | {\bf J}^2| \psi^- \rangle=0$, hence it provides
the realization of the null vector within two elementary systems.
Quantum probabilities for higher order angular momenta as well as for more complex operators as symmetric tensors can be obtained  following this approach.

In conclusion, our reformulation of QM based just on bits and classical physics,
provides a physical explanation to the most puzzling quantum features and specifically to entanglement. We hope that this conceptual understanding will be useful for the developing field of quantum information. 
Of course we stress that these results are not a complete reformulation and constitute only a promising starting point.
Despite we were able to reproduce results of the angular momentum in ordinary quantum theory, 
it should be evident that the present approach is far to be innocuous: Within this approach there is no room
to quantum observables displaying continuous outcomes in contrast to ordinary QM. This would imply that a tiny region of space can support only a finite amount of information. 
This consideration, descending naturally from the present approach, constitutes the common key ingredient of theories trying to match quantum theory with general relativity. We belive that the approach presented here may provide a rigorous  route towards the solution of this problem.

\begin{acknowledgements}
We thank O.\ Marag$\grave{\text{o}}$ and S.\ Portolan for helpful discussions and  comments.
\end{acknowledgements}

\newpage

\noindent Fig. 1 Scheme describing measurements for two particles with dicotomic outcomes and with null total angular momentum: (a) view of one of the two possible results when the two observers choose the same axis. (b) What happens when observer 2 choose a different rotation axis, according to the transformation properties of vectors under rotation.
\newline
\noindent

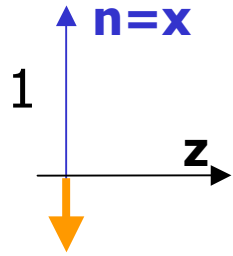
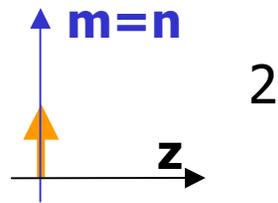
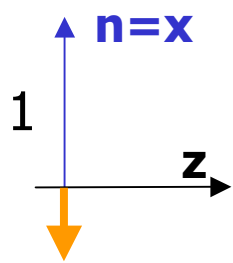
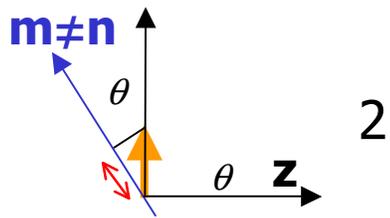
\end{document}